\documentclass[pteplogo]{ptephy_v2}

\preprintnumber{XXXX-XXXX} 
\usepackage{hyperref}

\usepackage{amsmath} 
\usepackage{amsthm} 
\usepackage{hyperref} 
\usepackage{graphics} 
\usepackage{booktabs}
\usepackage{array}
\usepackage{natbib}

\begin{document}

\title{Primordial Black Holes as a Factory of Axions: 
Extragalactic Photons from Axions}


\author[1,2]{Yongsoo Jho}
\author[1]{Tae-Geun Kim} 
\author[3,4]{Jong-Chul Park \thanks{jcpark@cnu.ac.kr; Correspondence}} 

\author[1,5]{Seong Chan Park \thanks{sc.park@yonsei.ac.kr; Correspondence}} 
\author[1]{Yeji Park}

\affil[1]{Department of Physics and IPAP, Yonsei University, Seoul 03722, Republic of Korea}
\affil[2]{Weizmann Institute of Science, Department of Particle Physics and Astrophysics, Rehovot, Israel 7610001}
\affil[3]{Department of Physics and Institute of Quantum Systems (IQS), Chungnam National University, Daejeon 34134, Republic of Korea}
\affil[4]{Particle Theory and Cosmology Group, Center for Theoretical Physics of the Universe, Institute for Basic Science (IBS),  Daejeon, 34126, Republic of Korea}
\affil[5]{Korea Institute for Advanced Study, Seoul 02455, Republic of Korea}

\begin{abstract}%
    Primordial black holes (PBHs) are significant sources of axions and axion-like particles (ALPs), provided their Hawking temperature exceeds the particles' masses. 
    Given the predominant decay of axions into photons, the enhanced photon spectrum they generate can be feasibly detected using sensitive detectors. 
    This paper introduces a novel methodology that elucidates the decay process for particles to traverse and decay over cosmological timescales.     
    Specifically, we derive estimations for the photon spectrum and flux, assuming a monochromatic mass spectrum and isotropic distribution for PBHs. 
    Encouragingly, forthcoming detectors like e-ASTROGAM are well positioned to capture this signal.
\end{abstract}

\subjectindex{xxxx, xxx}

\maketitle

\section{Introduction}
\label{sec:intro}

Primordial black holes (PBHs), thought to originate in the early universe~\cite{Hawking:1971ei}, might have been extensively generated during the inflationary period~\cite{Ivanov:1994pa, Garcia-Bellido:2017mdw, Pi:2017gih}, particularly in certain realistic Higgs inflation models~\cite{Hamada:2014wna, Hamada:2014iga, Ezquiaga:2017fvi, Cheong:2019vzl, Cheong:2022gfc}. 
PBHs are considered potential dark matter (DM) candidates~\cite{Bird:2016dcv, Carr:2016drx, Inomata:2017okj}.
PBHs could efficiently produce any new particles with a MeV-scale mass due to their relatively high Hawking temperature~\cite{Hawking:1975vcx}:
\begin{align}
T_H 
=\frac{\hbar c^3}{8\pi G k_B m_{\rm PBH}} \sim \left(\frac{10^{16}~{\rm g}}{m_{\rm PBH}}\right){\rm MeV}.
\end{align}
The axion is a prime example.

Axion~\cite{Peccei:1977hh, Peccei:1977ur} provides, arguably, the best-known solution to the strong-CP problem~\cite{Weinberg:1977ma, Wilczek:1977pj}. 
Axion-like particle (ALP) also arises in various theoretical models, providing solutions to some outstanding problems in particle physics and cosmology~\cite{Svrcek:2006yi, Lee:2014xua}.
It is intriguing to note that axion and ALP are likely to be in sub-MeV range for their roles~\cite{Kim:1979if, Shifman:1979if, Dine:1981rt, Zhitnitsky:1980tq, Morris:1984iz}.\footnote{See Refs.~\cite{Cheong:2022ikv, Hamaguchi:2021mmt} for a recent account on the axion quality problem.}
Various celestial bodies, including the Sun~\cite{DiLella:2000kr}, supernovae~\cite{Turner:1987by, Jaeckel:2017tud}, and neutron stars~\cite{Morris:1984iz}, have been proposed as potential sources of axions and ALPs~\cite{Choi:2020rgn}. 
This paper suggests PBHs as an alternative strong source. 
We are optimistic about the detection of excessive photon signals as ongoing observations and advancements in experimental techniques continue to progress~\cite{schonfelder1993instrument, Fermi-LAT:2009ihh, Abeysekara:2013tza, e-ASTROGAM:2017pxr}.

\begin{figure}[t]
\centering
\includegraphics[width=.6\linewidth]{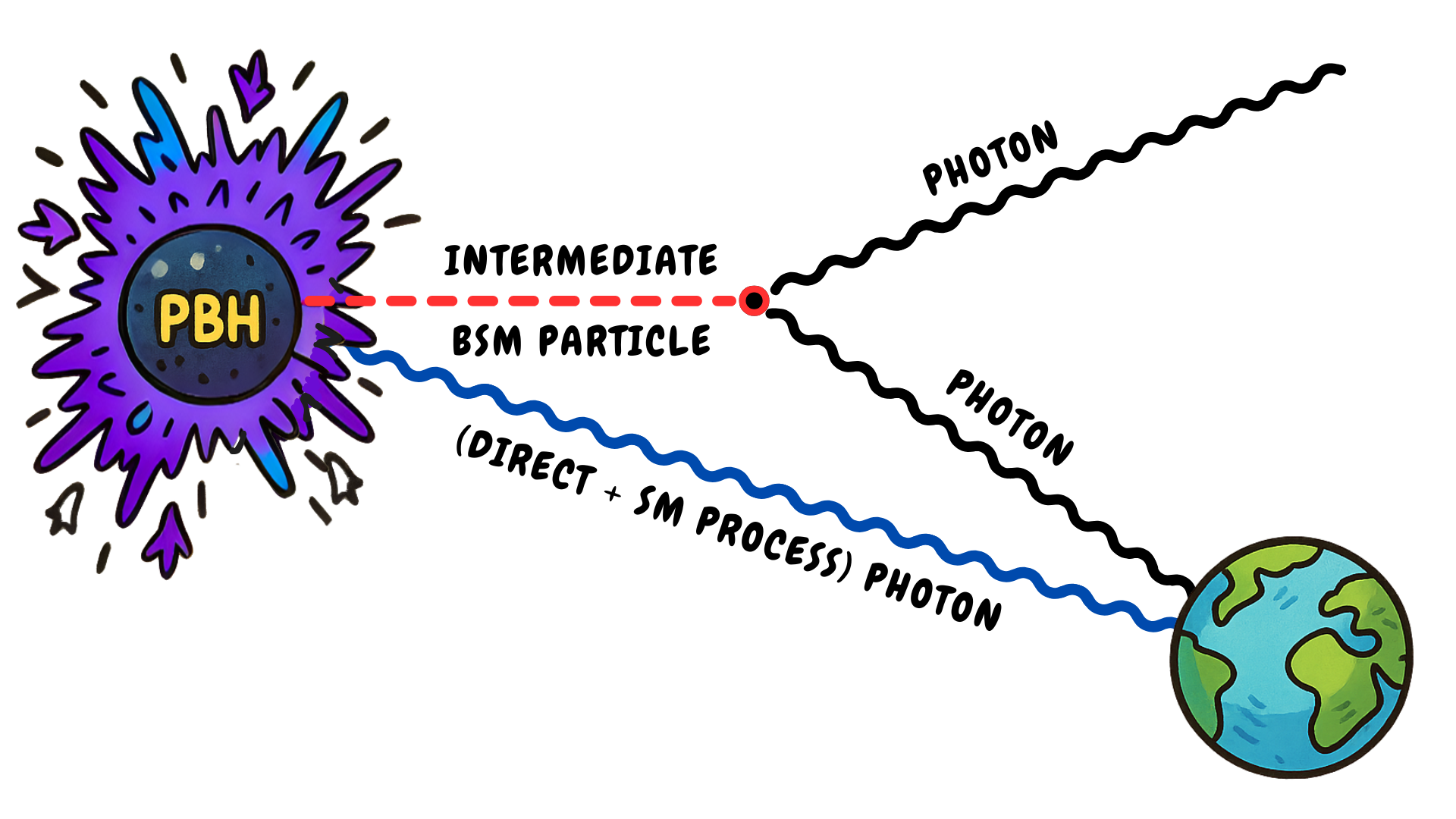}
\caption{A schematic picture of being considered. 
 The curly gray line shows the photon directly released and/or generated by standard model (SM) particles emitted from a PBH. 
 The curly black line represents photons produced by the decay of BSM particles emitted from the PBH. }
\label{fig:Process}
\end{figure}  

The schematic scenario we are considering in this paper is depicted in Fig.~\ref{fig:Process}. 
A PBH emits not only photons and neutrinos but all particle species with masses below the Hawking temperature, i.e., $m \lesssim T_H$~\cite{Page:1976df}. 
Gamma-ray photons are particularly interesting~\cite{Page:1976wx}, as they can propagate long distances without losing energy and eventually provide information about the source.\footnote{In principle, neutrinos also propagate to the Earth, but they are more useful in the ultra-high-energy regime~\cite{Rott:2014kfa}.} 
It is crucial to recognize that the photon spectrum comprises various contributions: (i) primary photons emitted directly from PBHs, characterized by a thermal spectrum with temperature $T_H$; (ii) secondary photons arising from the decay of hadrons; and finally, (iii) photons stemming from the decay of new particles like axions and ALPs. 
Each contribution plays a distinct role, necessitating strategies to effectively differentiate among these photon sources. 
We pay special attention to the third component, as it sensitively depends on the lifetime of the new particles.

The paper is structured as follows. 
In Section~\ref{sec:PBHs}, we set up our scenario with the properties of PBHs and the long-lived axion/ALP. 
In Section~\ref{sec:method}, we obtain photon fluxes by adding all components (i), (ii), and (iii). 
Importantly, we, for the first time, consider the effect of the decay elongation of axions over cosmological timescales. 
We present the currently available experimental bounds from the estimated photon signals in Section~\ref{sec:results} using the estimated photon fluxes. 
The summary is given in Section~\ref{sec:conclusion}.

\section{Hawking radiation of PBH}      \label{sec:PBHs}

The Hawking emission rate of a particle $X$ from a non-rotating PBH is~\cite{Hawking:1975vcx}
\begin{equation}  \label{eq:emission}
\frac{\text{d}^2 N_X}{\text{d}E\text{d}t} = \frac{g_X}{2\pi} \frac{\Gamma_X(E, m_{\text{PBH}})}{e^{E/T_{\text{PBH}}} - (-1)^{2s_X}}\,,
\end{equation}
where $g_X$ is the internal degrees of freedom of the emitted particle, $\Gamma_X$ is the greybody factors~\cite{Ida:2002ez, Ida:2005ax, Ida:2006tf}, and $s_X$ is the spin of the particle. 
In this analysis, we focus on non-rotating PBHs as the emission of angular momentum is efficient and the remaining PBHs are likely non-rotating~\cite{Page:1976df, Page:1976ki, Page:1977um}. 
We use  {\it BlackHawk}~\cite{Arbey:2019mbc} for the emission rate calculation.

Since PBHs are distributed on cosmological scales, the number density of PBHs changes with time. 
Taking this effect, the  photon flux is estimated as~\cite{Carr:2009jm}:
\begin{equation}
\begin{aligned}
        \dfrac{\text{d}F_{\gamma_0}}{\text{d}E_{\gamma_0}} = n_{\text{PBH}}(t_0)\displaystyle \int_{t_{\text{CMB}}}^{t_0} &\text{d}t\; (1+z(t)) \\
        &\times \frac{\text{d}^2N_\gamma}{\text{d}E_\gamma\text{d}t}\bigg|_{E_\gamma = (1+z(t))E_{\gamma_0}}\,,
\end{aligned}
\label{eq:dFdE}
\end{equation} 
where $t_0$ and $t_{\text{CMB}}$ respectively represent the current and CMB time. 
The parameter $n_{\text{PBH}}(t_0)= f_{\text{PBH}} \rho_{\text{DM}}/m_{\text{PBH}}$ represents the current number density of PBHs where $f_{\text{PBH}}$ is the PBH fraction of DM and $\rho_{\text{DM}}$ is the current DM energy density. 
The observed photon at present is denoted by $\gamma_0$, and $F_{\gamma_0}$ and $E_{\gamma_0}$ are the flux and energy of the photon, respectively. 
The redshift effect is taken into account by $z(t)$, and the total flux should be obtained by integrating over time after the CMB epoch as the photon would not propagate before.

\begin{figure}[t]
    \centering
    \includegraphics[width=0.6\linewidth]{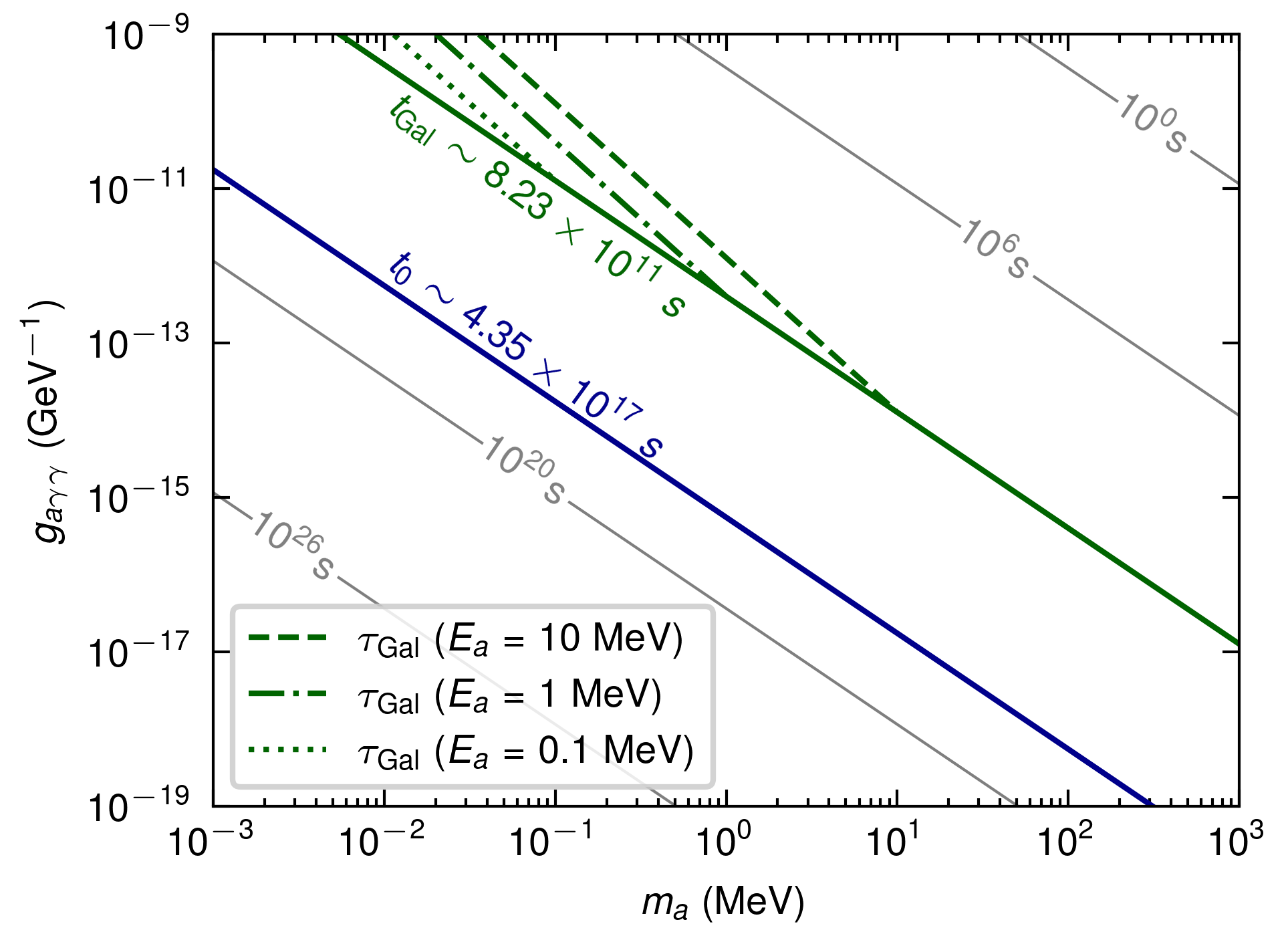}
    \caption{Lifetime of ALP in the rest frame. 
    The blue and green solid lines represent the cases where ALP lifetime equals the age of the universe and the GC-to-Earth distance, respectively.
    Given $E_a$ of emission ($E_a=0.1, 1, 10 \text{MeV}$), $\tau_{\text{Gal}} = t_{\text{Gal}} / \gamma(E_a)$ with the boost factor $\gamma(E_a) = E_a / m_a$.
    }
    \label{fig:tau}
\end{figure}

Since we focus on finding the axion and ALP signals, the cosmological timescale and distribution are all relevant. 
Indeed, in a large parameter space, an axion (ALP) has a cosmological scale lifetime and propagates long distance as depicted in Fig.~\ref{fig:tau}.
For long-lived axions (ALPs), the galactic component of gamma rays is negligible since most of them will decay far from the galactic center (GC) or Halo. 
Therefore, our focus is on the signals from the extragalactic component. 
This choice is also helpful to avoid complications arising in the GC region~\cite{Schiavone:2021imu, Li:2022mcf, Li:2022xqh}.

\medskip

\section{Photons from the axion decay}\label{sec:method}

When considering particles with cosmological-scale lifetimes, the effects of cosmic expansion become essential. 
Cosmic expansion induces two distinct effects: the dilution of particle number density and the redshift of particle energy. 
While most of previous studies have focused solely on the dilution effect, we emphasize that ``the energy redshift" is equally significant, as it affects the particle's decay rate through the modification of its boost factor.

Although our primary interest lies in axions and ALPs, it is worth noting that the analysis framework we develop here is more general. 
Indeed, our methodology can be applied to any long-lived particle that eventually decays into photons. 
Therefore, we first present our formalism for a generic particle $X$, before specializing to the case of axions. 

For a particle $X$ with mass $m_X$, the redshifted Lorentz boost factor is given by
\begin{align}
\gamma^{t;t_e}(E_X) = \frac{\mathcal{E}_X^{t;t_e}(E_X)}{ m_X}\,,
\end{align}
where $\mathcal{E}_X^{t;t_e}(E_X)$ is the redshifted energy at a time $t$ from the energy $E_X$ originally emitted from PBH at time $t_e$.

For long-lived particles, $t_e$ and $t$ can differ substantially, and in such a case the cosmological expansion can significantly affect the dynamics.
This manifests itself as a modification of the decay rate, leading to a redshifted decay rate at time $t$:
\begin{equation}
\Gamma_X^{t;t_e}(E_X) \equiv \frac{1}{\gamma^{t;t_e}(E_X)\tau_X}\,,
\end{equation}
where $\tau_X$ denotes the lifetime of the particle $X$ in its rest frame.

Taking the cumulative decay process from the earliest emission time $t_e^{\rm min}$ to $t$ and considering the survival probability function $P_{\rm surv}$, we obtain the differential number density of $X$ particles with energy $\tilde{E}_X$ decaying at time $t$ as

\begin{eqnarray}
\frac{\text{d}n_X^{\text{dec}}(\tilde{E}_X)}{\text{d}t} =  
    && \int_{t_e^\text{min}}^{t} \!\text{d}t_e \Big[n_{\mathrm{PBH}}(t_e)\left(\frac{1+z(t)}{1+z(t_e)}\right)^3 \\
    && \times E_X \frac{d^2 N_a}{d E_X d t_e} \mathcal{P}_{\text{decay}}(t; t_e, E_X) \Big]_{\mathcal{E}_X^{t;t_e}(E_X) = \tilde{E}_X}\,, \nonumber
\end{eqnarray}    

where $n_{\mathrm{PBH}}(t_e)$ is the number density of PBHs at emission time $t_e$, and $\mathcal{P}_{\text{decay}}(t; t_e, E_X) = \Gamma_X^{t:t_e}(E_X) P_\text{surv}(t;t_e,E_X)$ represents the ``decay probability density'' which gives the probability density function for a particle $X$ to decay at time $t$ after being emitted at $t_e$ with energy $E_X$. 
For a detailed derivation of the decay probability density, see Appendix \ref{time-varying}.

If we consider particle $X$ as an ALP, the redshifted decay rate can be written as follows
\begin{equation} \label{eq:decay_rate_ALP}
    \Gamma_a^{t;t_e}(E_a)= \frac{g_{a\gamma\gamma}^2 m_a^4}{64\pi \mathcal{E}_a^{t;t_e}(E_a)}\,.
\end{equation}

To obtain the final photon spectrum, we must properly account for the boosted decay kinematics ({\it BD kinematics}). 
The motion of the mother particle $X$ in the laboratory frame fundamentally affects the kinematics of its decay products, leading to significant modifications in their energy distribution~\cite{Kim:2015usa, Kim:2015gka}.\footnote{See Refs.~\cite{Agashe:2012bn, Agashe:2013eba} for the boosted collider physics signals.}

\begin{figure}[t]
    \centering
    \includegraphics[width=.6\linewidth]{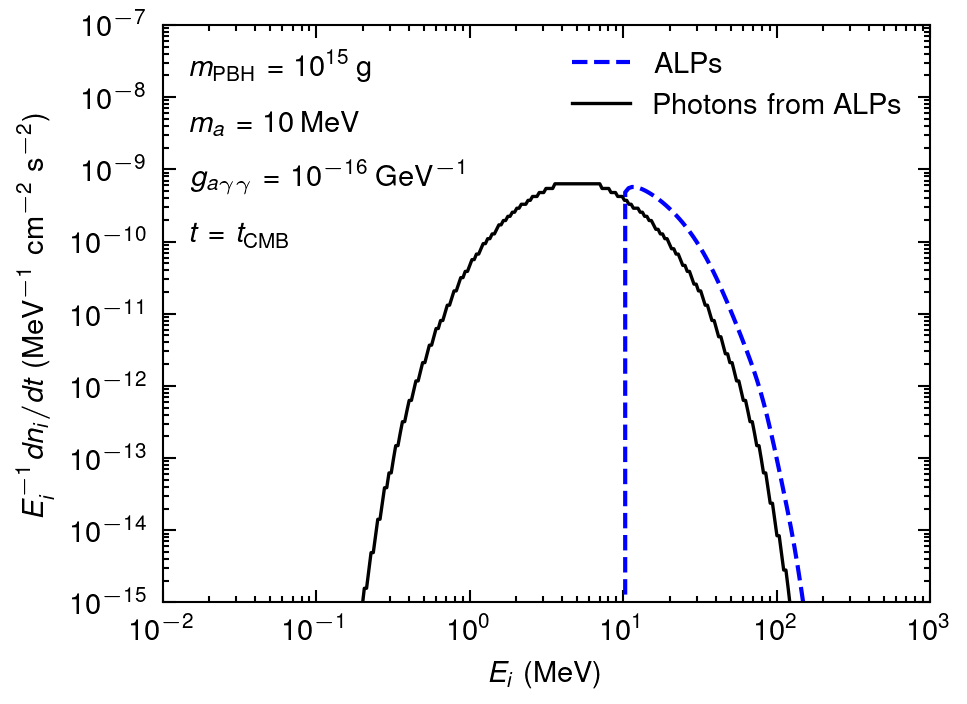}
    \caption{Photon spectrum from axion (ALP) spectrum. 
    The blue dashed line is the spectrum of decaying axion, while the solid black line is the corresponding photon spectrum.
    }
    \label{fig:boost}
\end{figure}

In Fig.~\ref{fig:boost}, we depict the axion (ALP) spectrum, shown with a blue dashed line, emitted during the cosmic microwave background (CMB) era, along with the photon spectrum resulting from the decay of axions (ALPs), represented by the black solid line. 
For this example, we consider the axion mass $m_a=10 \; \mathrm{MeV}$ and the PBH mass $m_{\rm PBH}=10^{15} \; \mathrm{g}$. 
The photon spectrum is characterized by a broader range skewed towards lower energies and features a significant peak at $E_\gamma=m_a/2$. 
The peak represents the energy of photons in the rest frame of their mother particle, which decays equally into two photons. 
The distribution of photons is broader at both lower and higher energies because of the Lorentz boost, which causes an energy spread.

\begin{figure}[t]
\centering
\includegraphics[width=.6\linewidth]{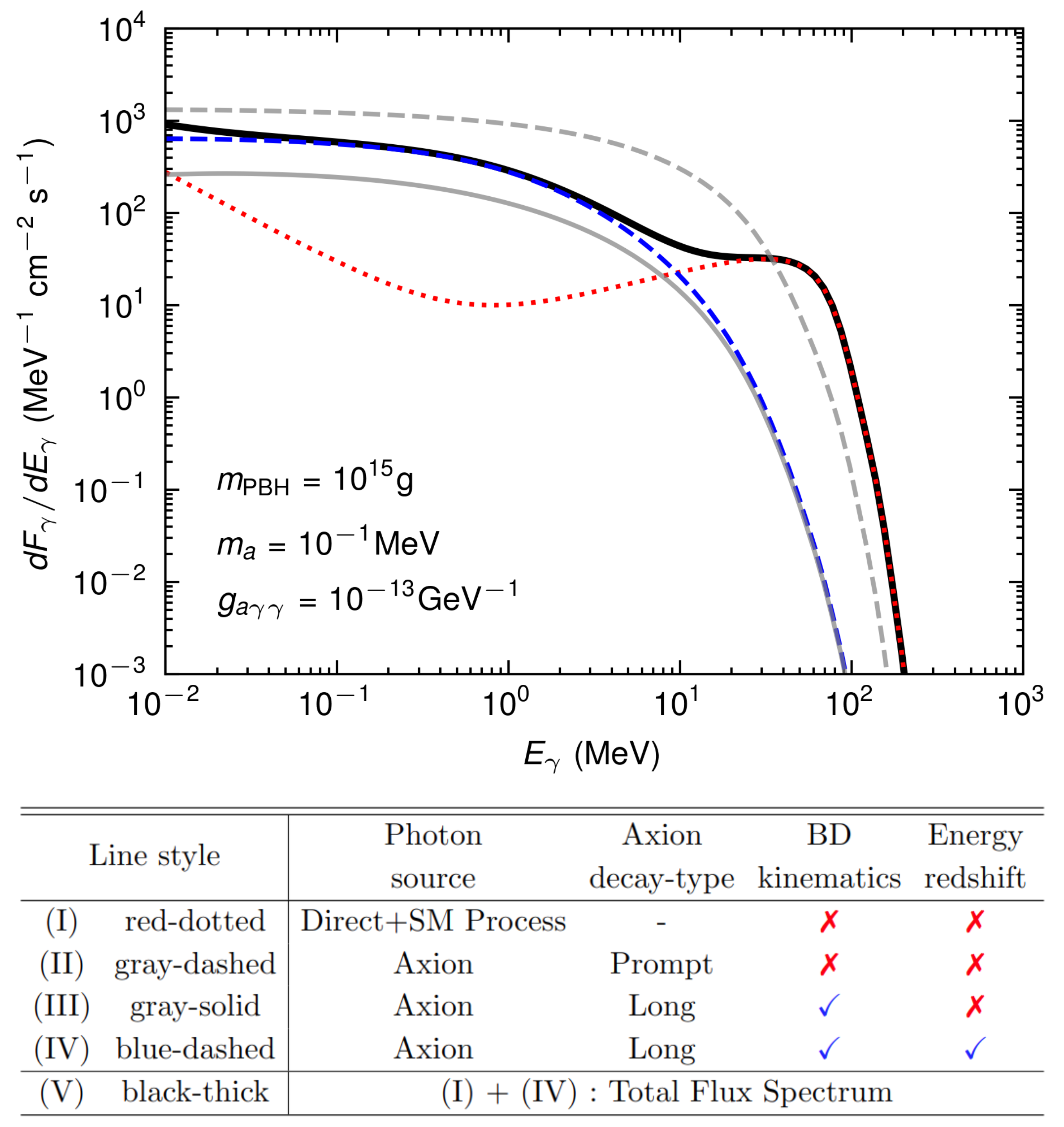}
\caption{
    Differential photon fluxes produced through various mechanisms. 
    The spectrum includes primary and secondary photons from Hawking radiation of SM degrees of freedom (red-dotted), photons from long-lived ALPs with cosmological-scale lifetime (blue-dashed), and their combined contribution (black-thick-solid). 
    For comparison, we have also included two reference cases: photons from promptly decaying ALPs (gray-dashed), which is a rather unrealistic assumption at the weak coupling regime considered here, and those calculated using conventional time-independent decay analysis (gray-solid). 
}
\label{fig:dFdE_new}
\end{figure}

Figure~\ref{fig:dFdE_new} demonstrates various contributions to the photon flux and highlights the importance of properly treating the axion decay process on cosmological timescales. 
The red-dotted line shows the combined contribution from primary and secondary photons produced through SM processes. 
The traditional approach that only considers the dilution effect of cosmic expansion (gray-solid line) significantly underestimates the complexity of the decay process. 
Our comprehensive analysis reveals that incorporating both effects of cosmic expansion, spatial dilution and energy redshift, along with the proper treatment of BD kinematics leads to distinctly different results (blue-dashed line). 
The total photon flux, combining both the SM contribution and our improved axion decay calculation, is shown by the black-thick-solid line. 
The enhancement in photon flux, particularly in the energy range $E_{\gamma} \sim (10^{-2}-10)~{\rm MeV}$, highlights the significance of these effects. 
For comparison, we also present the limiting case of prompt decay (gray-dashed line), which serves as an important theoretical benchmark to understand the complete decay dynamics.

\section{Observational sensitivity}   \label{sec:results}

\begin{figure}[t!]
    \centering
    \includegraphics[width=.6\linewidth]{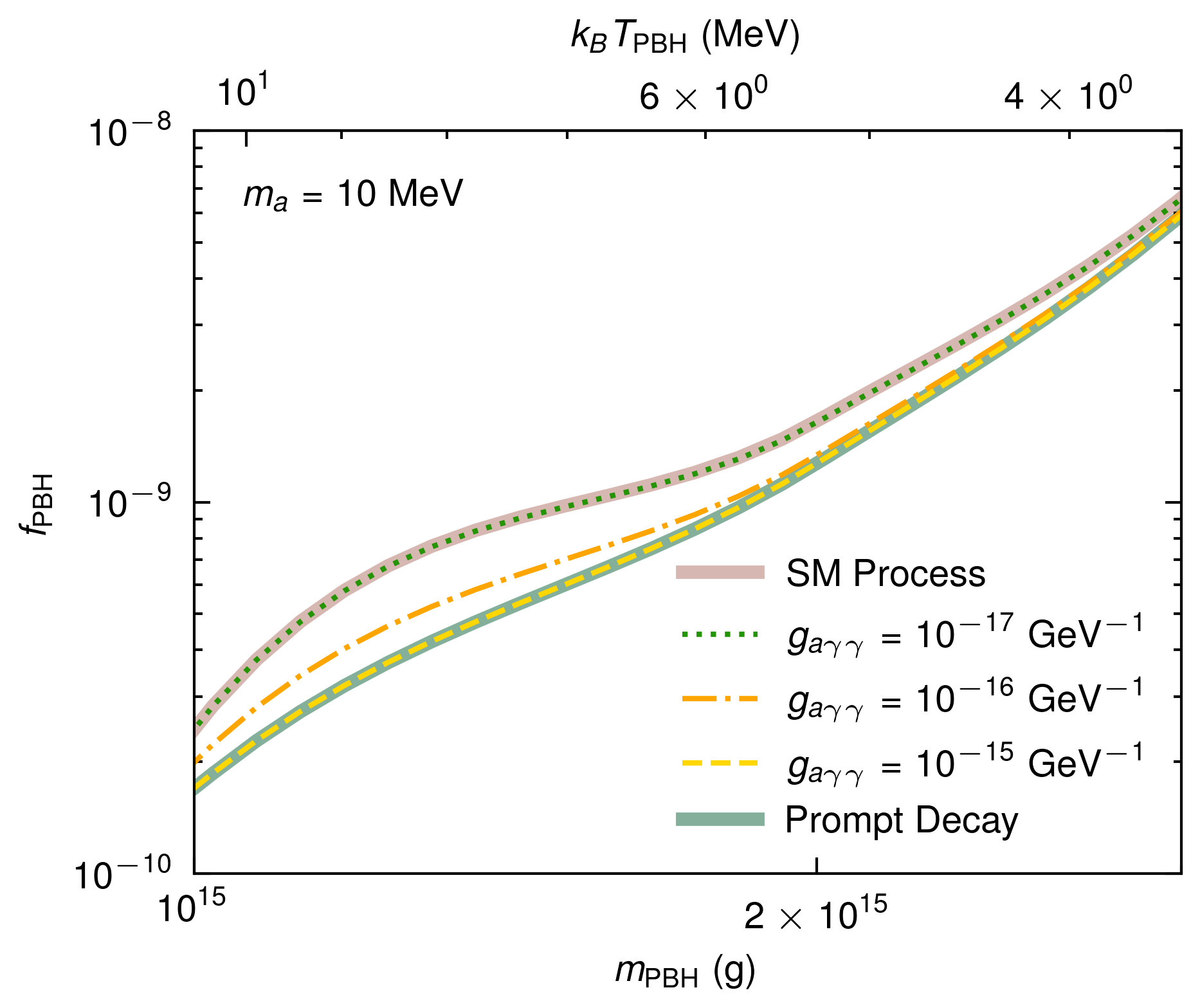}
    \caption{Expected upper limits on $f_{\text{PBH}}$ as a function of PBH mass $m_{\rm PBH}$ for three benchmark coupling constants $g_{a\gamma\gamma}$ with a fixed axion mass $m_a = 10 \; \text{MeV}$.
    }
    \label{fig:constraint_f_pbh}
\end{figure}

Finally, to study the future detection perspectives of the axions from PBHs, we take e-ASTROGRAM, a future space-based gamma-ray detection experiment in ESA, as a benchmark experiment. 
Indeed, e-ASTROGAM will improve sensitivity by one or two orders of magnitude in the energy range of $\mathcal{O}(\mathrm{MeV} - \mathrm{GeV})$ compared to existing ones~\cite{e-ASTROGAM:2017pxr}. 
Using the expected data given in Ref.~\cite{e-ASTROGAM:2017pxr}, we obtain the detection prospect of e-ASTROGAM: 
\begin{equation}
       I(E_{\gamma_0} \in (E_n, E_{n+1})) \equiv \int_{E_n}^{E_{n+1}} \frac{\text{d}F_{\gamma_0}}{\text{d}E_{\gamma_0}} \text{d}E_{\gamma_0}\,,
\end{equation}
where $\{E_n\}_{n=0}^{N}$ denotes a set of energy bins of e-ASTROGAM sensitivity. 
The limit of the abundance of PBH ($f_{\rm PBH}$) is obtained by the maximum value of the flux that is below the sensitivity for each bin. 
We compare the sensitivity limits for the SM only case and the case with the SM plus additional axion contribution. 
In Fig.~\ref{fig:constraint_f_pbh}, we present the expected upper limit on $f_{\rm PBH}$ for $m_{\rm PBH}\sim\mathcal{O}(10^{15}~{\rm g})$ and a fixed axion mass $m_a=10$ MeV. 
The limit approaches the SM-only limit when the coupling constant $g_{a\gamma\gamma}$ is small ($\lesssim 10^{-17} \;{\rm GeV}^{-1}$) so that it does not add observable photons. 
On the other hand, the bound approaches the prompt decay limit as the coupling constant increases ($\gtrsim 10^{-15} \; {\rm GeV}^{-1}$). 
Due to the significant contribution of axions, the bound of PBH can be improved around $m_{\rm PBH}=\mathcal{O}(10^{15}~{\rm g})$.

\begin{figure}[!t]
    \centering
    \includegraphics[width=.6\linewidth]{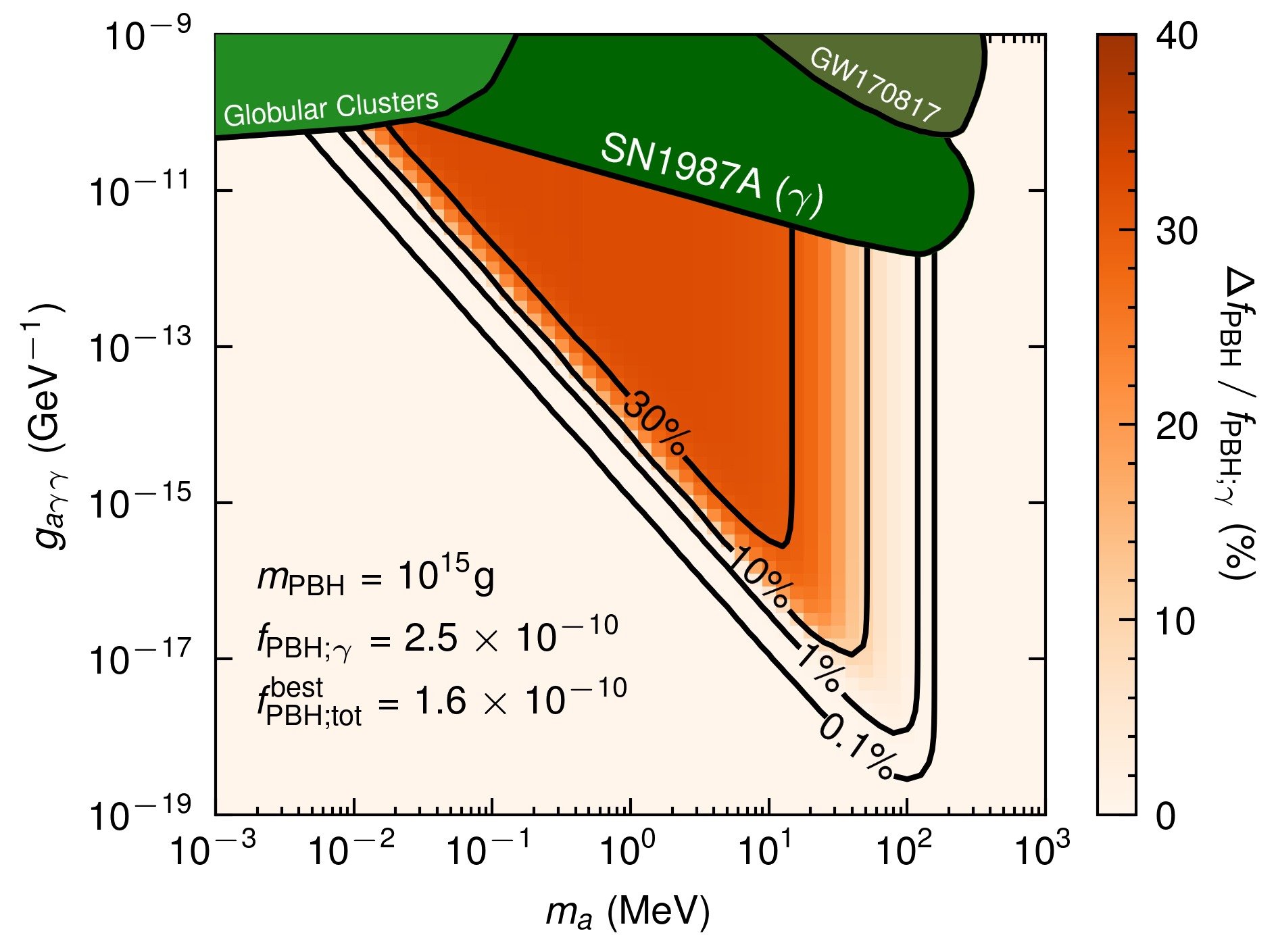}
    \caption{Deviation of expected limits on $f_{\rm PBH}$ with and without axion contributions.
    Black contours show the sensitivity growth due to the axion contributions.  
    The bounds from other astronomical sources are also included: Globular Clusters (light green)~\cite{Ayala:2014pea, Dolan:2022kul}, SN1987A (green)~\cite{Jaeckel:2017tud}, and GW170817 (dark green)~\cite{Diamond:2023cto}.
    }
    \label{fig:constraint_f_pbh_alp}
\end{figure}

In Fig.~\ref{fig:constraint_f_pbh_alp}, we highlight the difference between the case of SM alone and the case with axions. 
The PBH mass is set as the reference value $m_{\rm PBH}=10^{15}~{\rm g}$. 
We quantify the difference as follows,
\begin{align}
        \frac{\Delta f_{\rm PBH}}{f_{{\rm PBH;\,SM}}} \equiv \frac{f_{{\rm PBH;\,SM}} - f_{\rm PBH;\,SM+Axion}}{f_{{\rm PBH;\,SM}}}\,,
\end{align}
where $f_{\rm PBH;\,SM}$ and $f_{\rm PBH;\,SM+Axion}$ denote the PBH fraction bounds obtained in the SM-only case and the case with axion contributions, respectively.
Each black contour represents the deviation level of constraints on $f_{\rm PBH}$ due to the axion effects. 
The deviation can be larger than $30\%$ depending on the masses of axion and PBH, and the axion coupling constant. 
The green-shaded regions are also shown to compare with the limits from other astronomical sources such as supernovae, globular clusters, and gravitational waves. 

%

\section{Discussions}    \label{sec:conclusion}

The quantum nature of the black hole offers a completely different way of probing new particles beyond the standard model. 
Especially when the black hole is light, the Hawking temperature becomes high and opens a new window of producing light particles such as axion, ALP, and dark photon. 

We focused on the extragalactic component of gamma-ray signals as a probe of PBHs and checked the effects of the additional contribution from axions (and ALPs), especially when they are cosmologically long-lived. 
We carefully considered the cosmological effects in the decay process of such long-lived particles and found that the photon spectrum will be indeed modified, subject to being tested at future experiments such as e-ASTROGAM. 
We noted that the gamma-ray signal in $E_\gamma \in (1, 100)\; \mathrm{MeV}$ provides a particularly interesting opportunity to show the axion effects; therefore, e-ASTROGAM will have great potential to detect signals from PBHs as well as axions.

\section*{Acknowledgment}

We thank Bhaskar Dutta for bringing his work to our attention. 
The work is supported by the National Research Foundation of Korea (NRF) grant funded by the Korea government (MSIT) [NRF-2021R1A4A2001897 (JCP, SCP),  RS-2024-00356960 (JCP), RS-2023-00283129 and RS-2024-00340153 (SCP)].
The work of JCP is partially supported by IBS under the project code, IBS-R018-D1.
The work of SCP is also supported by Yonsei internal grant for Mega-science (2023-22-048).

\medskip
\noindent {\bf Note added.} During the completion of the paper, we became aware of a work~\cite{Agashe:2022phd} that focuses on the galactic contributions of PBHs covering different energy ranges.


\bibliographystyle{ptephy}
\bibliography{ref}

\clearpage

\appendix

\section{Notation and definitions of variables}

\begin{table}[ht]
\centering
\caption{
Notation and definitions adopted in this work. 
For a particle $X$ with mass $m_{X}$, we account for their energy evolution from $t_e$, at which corresponding initial energy is $E_{X}$, to $t$ (notated as $t;t_{e}$).}
\label{table:new_var}
\renewcommand{\arraystretch}{1.5} 
\small
\begin{tabular}{c@{\hspace{1em}}>{\raggedright\arraybackslash}p{7cm}>{\raggedright\arraybackslash}p{4.5cm}}
\toprule
Symbol & Meaning & Definition \\
\midrule
$t_0$ & Present time ($\sim 4.35 \times 10^{17}\text{s}$) & -- \\
$t_e$ & Emission time at PBHs & -- \\
$E_{X}$ & Initial energy of a particle $X$ & -- \\
\addlinespace[0.5em]
$f_{\text{PBH}}$ & The PBH fraction of DM abundance & $\displaystyle {\Omega_{\text{PBH}}}/{\Omega_{\text{DM}}}$ \\
\addlinespace[0.5em]
$n_{\text{PBH}}(t_0)$ & The number density of PBH at present & $\displaystyle {f_{\text{PBH}} \rho_{\text{DM}}}/{m_{\text{PBH}}}$ \\
\addlinespace[0.5em]
$\mathcal{D}\left(t_1; t_2\right)$ & Dilution factor by the expansion from $t_2$ to $t_1$ & $\displaystyle \left(\frac{1+z(t_1)}{1+z(t_2)}\right)^3$ \\
\addlinespace[0.5em]
$n_{\text{PBH}}(t)$ & The number density of PBH at the time $t$ & $\mathcal{D}\left(t; t_0\right)n_{\text{PBH}}(t_0)$ \\
\addlinespace[0.5em]
$\mathcal{E}_{X}^{t; t_e}\left(E_{X}\right)$ & Redshifted total energy & $\sqrt{m_{X}^2+\left(\frac{1+z(t)}{1+z\left(t_e\right)}\right)^2\left(E_{X}^2-m_{X}^2\right)}$ \\
\addlinespace[0.5em]
$\gamma_{X}^{t; t_e}\left(E_{X}\right)$ & Redshifted Lorentz factor & $ {\mathcal{E}_{X}^{t; t_e}\left(E_{X}\right) }/{m_{X}}$ \\
\addlinespace[0.5em]
$\Gamma_{X}^{t; t_e}\left(E_{X}\right)$ & Redshifted decay width in the laboratory frame & ${\Gamma_{X}^{\mathrm{rest}}}/{\gamma_{X}^{t; t_e}}$ \\
\addlinespace[0.5em]
$\tau_{X}^{t; t_e}\left(E_{X}\right)$ & Redshifted lifetime & $\gamma_{X}^{t; t_e} \tau_{X}^{\mathrm{rest}}$ \\
\addlinespace[0.5em]
$P_{\text {surv}}\left(t; t_e, E_{X}\right)$ & Survival probability & $\displaystyle \exp \left(-\int_{t_e}^t \Gamma_{X}^{t^{\prime}; t_e}\left(E_{X}\right) d t^{\prime}\right)$ \\
\addlinespace[0.5em]
$\mathcal{P}_{\text {decay}}\left(t; t_e, E_{X}\right)$ & Decay probability density & $\displaystyle \int_{t_e}^t \mathcal{P}_\text{decay}(t'; t_e, E_a)\text{d}t' = 1 - P_\text{surv}(t; t_e, E_a)$ \\
\bottomrule
\end{tabular}
\end{table}

\section{Decay process in an expanding universe} \label{time-varying}
\begin{figure}[t]
    \centering
    \includegraphics[width=.49\linewidth]{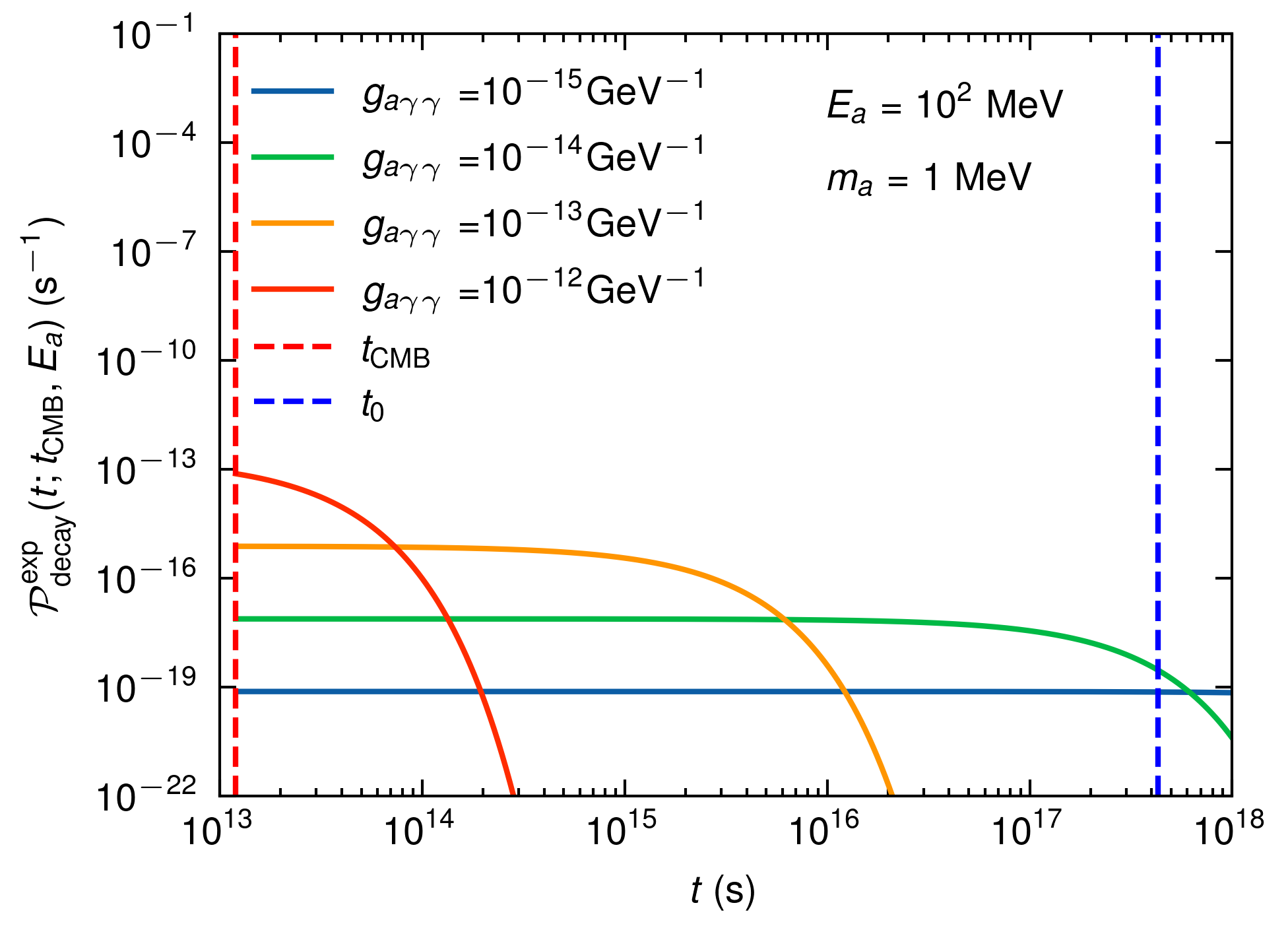}
    \includegraphics[width=.49\linewidth]{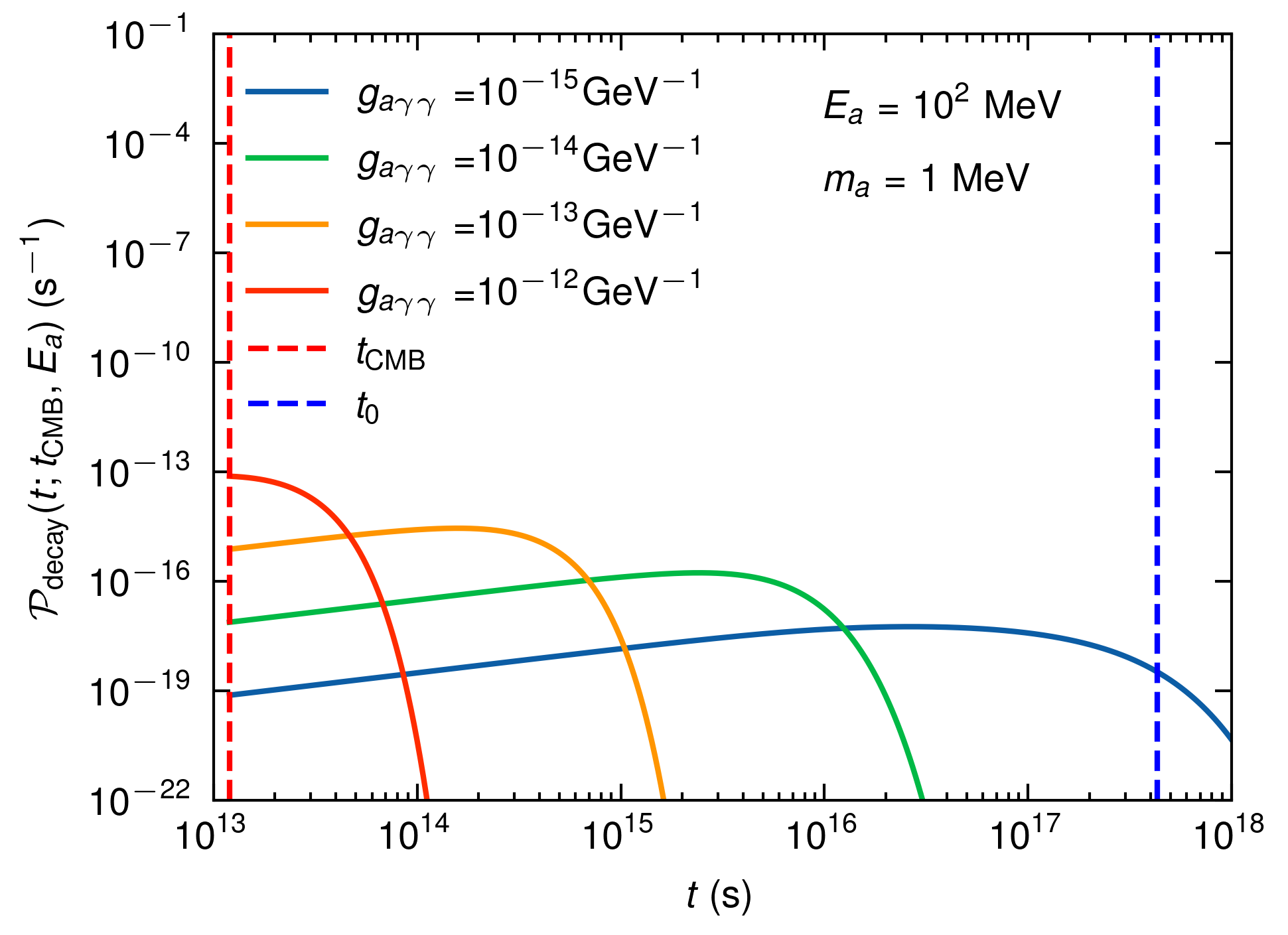}
    \caption{\textit{Decay probability density}: [Left] conventional exponential decay and [Right] decay in an expanding universe.}
    \label{fig:decaydensity}
\end{figure}

To determine the number of particle decays in an expanding universe, it is necessary to account for the temporal variation in the decay rate arising from the effects of cosmic expansion:
\begin{equation}
\begin{aligned}
    & \frac{\text{d}N_X}{\text{d}t} = -\Gamma_X^{t; t_e}(E_X) N_X \\
    \Longrightarrow~ & N_X(t; t_e, E_X) = N_X^e \cdot \exp \left(-\int_{t_e}^t \Gamma_X^{t'; t_e}(E_X)\,\text{d}t'\right)\,,
\label{eq:decay}
\end{aligned}
\end{equation}
where $N_X^e = N_X(t_e; t_e, E_X)$ denotes the initial number of particles $X$ produced at emission time $t_e$ with energy $E_X$.
In an expanding universe, the decay rate $\Gamma_X^{t; t_e}(E_X)$ evolves with time as the particle's energy undergoes a cosmic redshift. 
This time-dependent decay rate determines both the survival probability and the decay probability density at time $t$ for particles produced at time $t_e$ with energy $E_X$:
\begin{eqnarray}
    P_\text{surv}(t; t_e, E_X) &=& \exp \left(-\int_{t_e}^t \Gamma_X^{t'; t_e}(E_X)\,\text{d}t'\right), \\
    \mathcal{P}_\text{decay}(t; t_e, E_X) &=& \frac{1}{\tau_X^{t; t_e}(E_X)}P_\text{surv}(t; t_e, E_X)\,.
\end{eqnarray}
Here, $P_\text{surv}(t; t_e, E_X)$ represents the probability that particle $X$ survives until time $t$, with $\Gamma_X^{t'; t_e}(E_X)$ giving its instantaneous decay rate at time $t'$. 
The decay probability density $\mathcal{P}_\text{decay}(t; t_e, E_X)$ describes the probability density of decay events at time $t$, where $\tau_X^{t;t_e}(E_X)$ represents the instantaneous lifetime at time $t$, related to the decay rate by $\tau_X^{t;t_e}(E_X) = 1/\Gamma_X^{t; t_e}(E_X)$.
For comparison, in conventional exponential decay, these quantities take simpler forms:
\begin{eqnarray}
P_\text{surv}^\text{exp}(t; t_e, E_X) &=& \exp \left(-\frac{\Gamma_X(E_X)}{\gamma}(t-t_e)\right), \\
\mathcal{P}_\text{decay}^\text{exp}(t; t_e, E_X) &=& \frac{1}{\gamma\tau_X(E_X)} P_\text{surv}^\text{exp}(t; t_e, E_X)\,,
\end{eqnarray}
where $\Gamma_X(E_X)$ represents the constant decay rate with energy $E_X$, $\tau_X(E_X)$ is the particle's rest-frame lifetime, and $\gamma$ denotes the Lorentz factor. These expressions differ fundamentally from the expanding universe case, where the time integral in the exponent captures the evolving decay rate.
This distinction leads to markedly different behavior in the decay probability density, as illustrated in Fig.~\ref{fig:decaydensity}. For exponential decay (left panel), the decay probability density decreases monotonically with time. In contrast, decay in an expanding universe (right panel) exhibits an initial period where the survival probability remains near unity ($\approx 1$). As cosmic expansion progressively reduces the particle's energy, its lifetime shortens, causing the decay probability density to increase over an extended period.

\begin{table}[ht]
\centering
\caption{Comparison between survival analysis and decay in an expanding universe}
\footnotesize
\begin{tabular}{>{\centering\arraybackslash}p{3.2cm} >{\centering\arraybackslash}p{3cm} | >{\centering\arraybackslash}p{3.5cm} >{\centering\arraybackslash}p{3.5cm}}
\toprule
\multicolumn{2}{c|}{Survival analysis} & \multicolumn{2}{c}{Decay in expanding universe} \\[0.5ex]
Term & Notation & Term & Notation \\
\midrule 
Survival function & $S(t) = \mathbb{P}[X > t]$ & Survival probability & $P_\text{surv}(t; t_e, E_X)$ \\[1em]
Hazard function & $h(t) = -\frac{\text{d}}{\text{d}t} [\log S(t)]$ & Decay rate & $\Gamma_X^{t; t_e}(E_X)$ \\[0.5em]
Failure density function & $\textstyle \int_0^t f(u)\text{d}u = 1-S(t)$ & Decay probability density & $\textstyle \int_{t_e}^t \mathcal{P}_\text{decay}(t'; t_e, E_X)\text{d}t' = 1 - P_\text{surv}(t; t_e, E_X)$ \\
\bottomrule
\end{tabular}
\label{tab:surv}
\end{table}
To describe particle decay in an expanding universe, it is essential to establish a framework that extends beyond the limitations of conventional decay processes. In this work, we adopt the mathematical framework of \textit{Survival Analysis} (SA) as outlined in Ref.~\cite{kleinbaum_klein_2020}, which provides powerful tools for analyzing time-dependent phenomena.

Table~\ref{tab:surv} highlights the correspondence between key concepts from survival analysis and particle decay in an expanding universe. The survival function, $S(t)$, represents the probability that a particle survives beyond time $t$. As a complementary cumulative distribution function (tail function), $S(t)$ must satisfy the following two conditions:
\begin{itemize}
    \item $S(0) = 1$,
    \item $S(t)$ is a monotonically decreasing function.
\end{itemize}
Similarly, the survival probability $P_\text{surv}(t; t_e, E_X)$ satisfies the following conditions:
\begin{eqnarray}
    P_\text{surv}(t_e; t_e, E_X) &=& \exp\left(-\int_{t_e}^{t_e} \Gamma_X^{t'; t_e}(E_X)\,\text{d}t'\right) = 1\,, \\
    \frac{\text{d}\log{P_\text{surv}}}{\text{d}t} &=& -\Gamma_X^{t; t_e}(E_X) = -\frac{1}{\tau_X^{t; t_e}(E_X)} < 0\,.
\end{eqnarray}
In survival analysis, the failure rate measures the frequency of failures within a specified time interval and is defined as:
\begin{equation}
    \lambda(t) = \frac{S(t_1) - S(t_2)}{(t_2 - t_1)S(t_1)}\,.
\end{equation}
The hazard function, $h(t)$, represents the instantaneous failure rate and is defined as:
\begin{equation}
    h(t) = \lim_{\Delta t \to 0} \frac{S(t) - S(t+\Delta t)}{\Delta t \cdot S(t)} = -\frac{1}{S(t)}\frac{\text{d}S(t)}{\text{d}t}\,.
\end{equation}
In the context of particle decay in an expanding universe, the hazard function corresponds directly to the decay rate $\Gamma_X^{t; t_e}$. 
The failure density function, $f(t)$, further complements the description of the continuous hazard process through the relations:
\begin{equation}
    1 - S(t) = \int_0^t f(\tau)\,\text{d}\tau ~\Longrightarrow~ h(t) = \frac{f(t)}{S(t)}\,.
\end{equation}
The decay probability density follows a similar mathematical structure:
\begin{eqnarray}
    \mathcal{P}_\text{decay}(t; t_e, E_X) &=& \Gamma_X^{t; t_e} \cdot P_\text{surv}(t; t_e, E_X)\,, \\
    \int_{t_e}^t \mathcal{P}_\text{decay}(t'; t_e, E_X)\,\text{d}t' &=& 1 - P_\text{surv}(t; t_e, E_X)\,.
\end{eqnarray}

\end{document}